\documentclass[aps,prb,twocolumn,superscriptaddress,showpacs]{revtex4}
\usepackage{eurosym}
\usepackage{amsfonts}
\usepackage{amssymb}
\usepackage{amsmath}
\usepackage{graphicx}
\usepackage{color}
\begin{document}

\title{CuSe-based layered compound Bi$_{2}$YO$_{4}$Cu$_{2}$Se$_{2}$ as a quasi-two-dimensional metal}

\author{S. G. Tan}
\email[The authors contributed equally to this work.]{}

\affiliation{Key Laboratory of Materials Physics, Institute of Solid State Physics, Chinese Academy of Sciences, Hefei 230031, People's Republic of China}

\author{D. F. Shao}
\email[The authors contributed equally to this work.]{}

\affiliation{Key Laboratory of Materials Physics, Institute of Solid State Physics, Chinese Academy of Sciences, Hefei 230031, People's Republic of China}

\author{W. J. Lu}
\affiliation{Key Laboratory of Materials Physics, Institute of Solid State Physics, Chinese Academy of Sciences, Hefei 230031, People's Republic of China}

\author{B. Yuan}
\affiliation{Key Laboratory of Materials Physics, Institute of Solid State Physics, Chinese Academy of Sciences, Hefei 230031, People's Republic of China}

\author{Y. Liu}
\affiliation{Key Laboratory of Materials Physics, Institute of Solid State Physics, Chinese Academy of Sciences, Hefei 230031, People's Republic of China}

\author{J. Yang}
\affiliation{Key Laboratory of Materials Physics, Institute of Solid State Physics, Chinese Academy of Sciences, Hefei 230031, People's Republic of China}

\author{W. H. Song}
\affiliation{Key Laboratory of Materials Physics, Institute of Solid State Physics, Chinese Academy of Sciences, Hefei 230031, People's Republic of China}

\author{Hechang Lei}
\email[Corresponding author: ]{hchlei@issp.ac.cn}
\affiliation{Key Laboratory of Materials Physics, Institute of Solid State Physics, Chinese Academy of Sciences, Hefei 230031, People's Republic of China}

\author{Y. P. Sun}
\email[Corresponding author: ]{ypsun@issp.ac.cn}
\affiliation{Key Laboratory of Materials Physics, Institute of Solid State Physics, Chinese Academy of Sciences, Hefei 230031, People's Republic of China}
\affiliation{High Magnetic Field Laboratory, Chinese Academy of Sciences, Hefei 230031, People's Republic of China}
\affiliation{University of Science and Technology of China, Hefei 230026, People's Republic of China}

\makeatletter


\begin{abstract}
We have investigated the physical properties of a new layered oxyselenide
{{Bi$_{2}$YO$_{4}$Cu$_{2}$Se$_{2}$}}, which crystallizes
in an unusual intergrowth structure with {\normalsize{Cu$_{2}$Se$_{2}$}}
and{{ Bi$_{2}$YO$_{4}$}} layers. Electric transport measurement
indicates that {{Bi$_{2}$YO$_{4}$Cu$_{2}$Se$_{2}$}}
behaves metallic. Thermal transport and Hall measurements show that
the type of the carriers is hole-like and it may be a potential thermoelectric
material at high temperatures. First principle calculations are in
agreement with experimental results and show that {{Bi$_{2}$YO$_{4}$Cu$_{2}$Se$_{2}$}}
is a quasi-2D metal. Further theoretical investigation suggests the ground states of the {{Bi$_{2}$YO$_{4}$Cu$_{2}$Se$_{2}$}}-type can be tuned by designing the blocking layers, which will enrich the physical properties of these compounds.

\end{abstract}
\maketitle

\section{INTRODUCTION}

Quasi-two-dimensional (quasi-2D) oxide compounds of transition-metal
elements have been studied extensively because of their exotic and
superior electrical and magnetic properties, such as high-temperature
superconducting cuprates,\cite{Wong-ng-WK-1988} manganites with colossal
magnetoresistance,\cite{Kimura-T-2000} and cobaltite-based thermoelectric
materials, \cite{Maignan-A-2002} etc.. In contrast, oxychalcogenides,
which also tend to adopt layered structure due to the different sizes
and coordination requirements of oxygen and the heavier chalcogenide
anions, are a relatively under-investigated class of solid-state compounds.\cite{Clarke-SJ-2008}
Among known layered TMCh-based (TM = Cu, Ag; Ch =
S, Se, Te) oxychalcogenides, there are mainly two structural types.\cite{Clarke-SJ-2008} One structural type is
representative by the LnOTMCh series (Ln = lanthanide),
which can be regard as the filled PbFCl-type structure and is isostructural
to ZrSiCuAs as well as iron-based superconductors LnOFePn (Pn = As,
P).\cite{Kamihara-Y-2006,Kamihara-Y-2008} These compounds composed
of alternating Ln$_{2}$O$_{2}$ fluorite layers and TM$_{2}$Ch$_{2}$
antifluorite layers along $c$ axis have been studied intensively
because of their superior and novel optical and transport properties
originating from TM$_{2}$Ch$_{2}$ layers.\cite{Palazzi-M-1980,Palazzi-M-1981,Ueda-K-2000,Hiramatsu-H-2005}
Another important structural type is representative by the
oxide antimonide Sr$_{2}$Mn$_{3}$Sb$_{2}$O$_{2}$\cite{Brechtel-E-1979}
(which may conveniently be formulated as Sr$_{2}$MnO$_{2}$Mn$_{2}$Sb$_{2}$)
and related oxide pnictides.\cite{Enjalran-M-2000,Ozawa-TC-2001}
Later on, this structural type is also found in Sr$_{2}$ZnO$_{2}$Cu$_{2}$S$_{2}$.\cite{Zhu-WJ-1997}
In Sr$_{2}$ZnO$_{2}$Cu$_{2}$S$_{2}$, the chalcogenide layer {[}Cu$_{2}$S$_{2}${]}$^{2-}$
is similar to that in LaOCuS whereas the {[}Sr$_{2}$ZnO$_{2}${]}$^{2+}$
is not fluorite type but perovskite type. Several oxychalcogenides
with this crystal structure have been reported.\cite{Zhu-WJ-1997,Zhu-WJ-1997-2,Otzschi-K-1999,Adamson-P-2012,Smura-CF-2011,Jin-SF-2012}

\begin{table}

\caption{\label{Table1-konwn-CuSe} Some typical CuCh-based compounds and their
ground states.}

\begin{tabular}{ccc}
\hline
\hline
Compound & & Ground state\tabularnewline
\hline
KCu$_{2}$Se$_{2}$ &~~~~~~& Metallic\cite{Tiedje-O-2003}\tabularnewline
BaCu$_{2}$Se$_{2}$ &~& Semiconducting\cite{Krishnapriyan-2013}\tabularnewline
LaOCuS(Se) &~& Semiconducting\cite{Hiramatsu-H-2008}\tabularnewline
BaFCuS(Se) &~& Semiconducting\cite{Yanagi-H-2006}\tabularnewline
BiOCuS(Se) &~& Semiconducting\cite{Hiramatsu-H-2008}\tabularnewline
YOCuSe &~& Semiconducting\cite{Ueda-K-2007}\tabularnewline
HgOCuSe &~& Metallic\cite{Kim-GC-2012}\tabularnewline
Sr$_{2}$MnO$_{2}$Cu$_{2}$Se$_{2}$ &~& Semiconducting\cite{Jin-SF-2012}\tabularnewline
Sr$_{3}$Sc$_{2}$O$_{5}$Cu$_{2}$S$_{2}$ &~& Semiconducting\cite{Liu-ML-2007,Scanlon-DO-2009}\tabularnewline
Bi$_{2}$YO$_{4}$Cu$_{2}$ Se$_{2}$ &~& Metallic\tabularnewline
\hline
\hline
\end{tabular}

\end{table}

As listed in Table \ref{Table1-konwn-CuSe}, most of the reported
CuCh-based layered compounds are semiconductors.\cite{Jin-SF-2012,Krishnapriyan-2013,Hiramatsu-H-2008,Yanagi-H-2006,Liu-ML-2007,Scanlon-DO-2009} But
it is well-established \cite{Vajenine-GV-1996} that the copper chalcogenide
layers present in the oxysulfides can readily accept holes in the
antibonding states at the top of a valence band that is composed of
well-mixed Cu-$3d$ and S-$3p$ orbitals, and these holes are very
mobile. For example, La$_{1-x}$Sr$_{x}$OCuS is a rare example of
a transparent $p$-type semiconductor \cite{Takano-Y-1995} in which
the Cu-$3d$/S-$3p$ valence band is doped with holes; the band gap
insulator Sr$_{2}$ZnO$_{2}$Cu$_{2}$S$_{2}$ can also be doped to
produce a metallic Na$_{x}$Sr$_{2-x}$ZnO$_{2}$Cu$_{2}$S$_{2}$
with $p$-type carriers with $x\sim0.1$,\cite{Ueda-K-2001} and there
is an evidence for the very high mobility of $p$-type carriers in the
related Sr$_{3}$Sc$_{2}$O$_{5}$Cu$_{2}$S$_{2}$ with Cu or Sr
deficiencies.\cite{Liu-ML-2007,Scanlon-DO-2009} On the other hand,
HgOCuSe \cite{Kim-GC-2012} isostructural to LaOCuS is a metal  and
KCu$_{2}$Se$_{2}$ \cite{Tiedje-O-2003} in which La$_{2}$O$_{2}$ layer is replaced by
alkali metal K also exhibit metallic behavior.
Therefore, the physical properties of CuCh-based layered oxychalcogenides
are not only determined by the intralayer interaction of CuCh layer
alone but also by the interlayer interactions.

Recently, a new series of layered oxychalcogenides Bi$_{2}$LnO$_{4}$Cu$_{2}$Se$_{2}$
(Y, Gd, Sm, Nd, and La) were synthesized.\cite{Evans-J-S-O-2002}
The metallic behavior is stated for the compound Bi$_{2}$YO$_{4}$Cu$_{2}$Se$_{2}$
which is very rare in the CuCh-based oxychalcogenides. In this work,
we systematically studied the magnetic property, electric transport, heat capacity, thermoelectric
properties and the electronic structure of the layered oxyselenide
Bi$_{2}$YO$_{4}$Cu$_{2}$Se$_{2}$. These results indicate the quasi-2D metallic ground
state in this compound. We also found it may be a potential thermoelectric
material at high temperatures. Furthermore, we systematically studied
the doping effects of such system in theory. The results suggest that the ground state of the Bi$_{2}$YO$_{4}$Cu$_{2}$Se$_{2}$-type compounds can be tuned by designing the blocking layers.

\section{EXPERIMENT SECTION}

Bi$_{2}$YO$_{4}$Cu$_{2}$Se$_{2}$ sample was prepared by reacting a stoichiometric mixture
of Bi$_{2}$O$_{3}$, Y$_{2}$O$_{3}$, Y, Cu, and Se. The raw materials
were mixed and ground thoroughly in an agate pestle and mortar, and
then the mixtures were pressed into pellets under 12 MPa. The pellets
were placed into dried alumina crucibles and sealed under vacuum ($<10^{-4}$
Pa) in the silica tubes which had been baked in drybox for 1-2 h at
$150\,^{\circ}\mathrm{C}$. The ampoules were heated to $830\,^{\circ}\mathrm{C}$
with $1\,^{\circ}\mathrm{C}/\mathrm{min}$ and maintained at this temperature
for 24 h. Finally the furnace is shut down and cools to room temperature
naturally. The obtained samples were reground, pelletized, and heated
for another 24 hours at 830 $^{\circ}$C followed by furnace cooling.

The X-ray powder diffraction patterns were recorded at room temperature
on a Panalytical diffractometer (X\textquoteright{}Pert PRO MRD) with
Cu-$K_\alpha$ radiation and a graphite monochromator
in a reflection mode. Structural refinement
of powder Bi$_{2}$YO$_{4}$Cu$_{2}$Se$_{2}$ sample was carried
out by using Rietica software.\cite{Rietica} The average stoichiometry of
a Bi$_{2}$YO$_{4}$Cu$_{2}$Se$_{2}$ polycrystalline was determined
by examination of multiple points using an energy-dispersive X-ray
spectroscopy (EDX) in a JEOL JSM-6500 scanning electron microscope. The XPS data were taken on an AXIS-Ultra instrument from Kratos using
monochromatic Al K$_\alpha$ radiation (225 W, 15 mA, 15 kV) and
low-energy electron flooding for charge compensation. The magnetic susceptibility was measured by a superconducting quantum interference device magnetometer (Quantum Design MPMS). Electron paramagnetic resonance (EPR) spectra were recorded using an X-band Bruker EMX plus 10/12 cw spectrometer operating at 9.4 GHz equipped with an Oxford continuous flow cryostat within the temperature range of 2 $\sim$ 300 K. The electrical resistivity, thermal transport property, specific heat and Hall measurements were measured using a
Quantum Design Physical Properties Measurement System (PPMS).

Electronic structure was obtained from first-principles density functional
theory (DFT) in the generalized gradient approximation (GGA) according
to the Perdew-Burke-Ernzerhof.\cite{PBE-1996} The QUANTUM-ESPRESSO
package\cite{QE-2009} was used with ultrasoft pseudopotential from
GBRV pseudopotential library.\cite{GBRV-2014} The discussion about
the Coulomb correlation of $3d$ electrons in Bi$_{2}$YO$_{4}$Cu$_{2}$Se$_{2}$
was carried out using the GGA+$U$ approach in the version introduced
by Anisimov $et$ $al.$\cite{Anisimov-VI-1993} with an approximate self-interaction
correction implemented in the rotationally invariant way according
to Liechtenstein $et$ $al.$.\cite{Liechtenstein-AI-1995} We tested different
values of the screened Coulomb parameter $U$. The exchange parameter
$J$ was fixed to $\sim0.1U$. The energy cutoff for the plane-wave basis
set was 40 Ry. Brillouin zone sampling is performed on the Monkhorst-Pack
(MP) mesh \cite{Monkhorst-HJ-1976} of $16 \times 16 \times 16$.

\section{RESULTS AND DISCUSSION}

\subsection{Crystal Structure, Composition, and Valence State. }

\begin{figure}
\includegraphics[width=0.8\columnwidth]{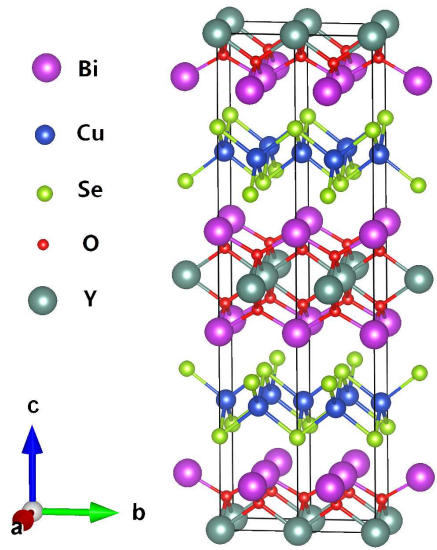}\caption{\label{fig1-structure}The crystal structure of Bi$_{2}$YO$_{4}$Cu$_{2}$
Se$_{2}$. }

\end{figure}
\begin{figure}
\includegraphics[width=0.95\columnwidth]{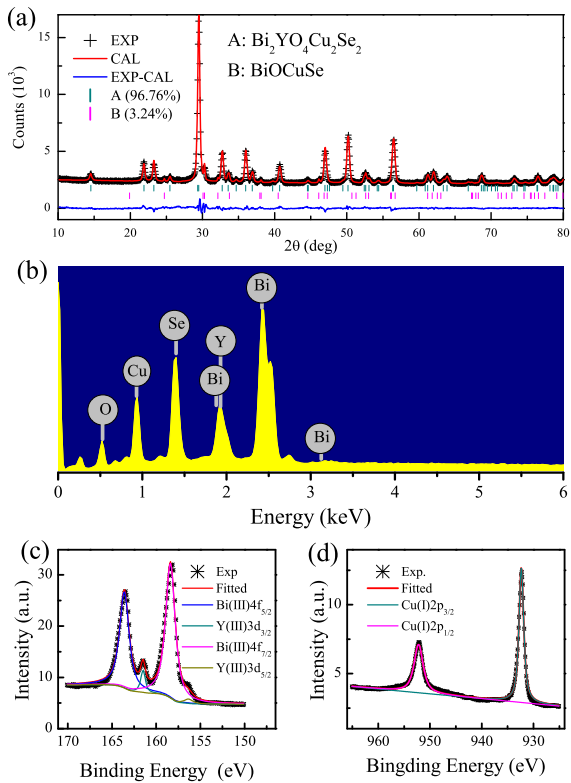}\caption{\label{fig2-XRD-EDS}(a) Result of Rietveld refinement against powder
XRD diffraction data for Bi$_{2}$YO$_{4}$Cu$_{2}$Se$_{2}$; (b) is the EDX spectrum of the sample; (c) and (d) are the XPS spectra
of Bi, Y and Cu.}

\end{figure}

The structure of Bi$_{2}$YO$_{4}$Cu$_{2}$Se$_{2}$ is shown in
Fig. \ref{fig1-structure}, which can be described as stacking of
edge-shared CuSe$_{4}$ tetrahedron layers with Bi$_{2}$YO$_{4}$
layers alternatively along $c$ axis. Fig. \ref{fig2-XRD-EDS} (a)
shows the XRD pattern and the result of Rietveld refinement. The structural
parameters obtained from the Rietveld refinements are listed in Tables
\ref{Table2-atomic-coordinates} and \ref{Table3-Structural-parameters}.
Almost all the reflections can be indexed in the space group of I4/mmm.
The refined lattice parameters of Bi$_{2}$YO$_{4}$Cu$_{2}$Se$_{2}$
are $a=3.8628(3)$ \r{A} and $c=24.322(2)$ \r{A}, which is consistent
with the previous reported results for Bi$_{2}$YO$_{4}$Cu$_{2}$Se$_{2}$.\cite{Evans-J-S-O-2002}
The other weak reflections originate from the BiOCuSe impurity, and
the result of the two-phase fitting reveals that the weight percentage
of BiOCuSe is about 3.24\%, which can be neglected to our study on
Bi$_{2}$YO$_{4}$Cu$_{2}$Se$_{2}$. The EDX spectrum (Fig. \ref{fig2-XRD-EDS}
(b)) of polycrystal confirms the presence of Bi, Y, O, Cu, and Se.
The average atomic ratios determined from EDX are Bi: Y: Cu: Se =
1.99(4) : 1.00(2): 1.99(1) : 1.81(5) when setting the content of Y
as 1. It is consistent with the formula of Bi$_{2}$YO$_{4}$Cu$_{2}$Se$_{2}$ except a small amount of Se deficiency, which may be caused by the easily volatilization of Se.
In order to determine the valence state of Cu, Bi, and Y, the XPS
spectrum was measured for the Bi$_{2}$YO$_{4}$Cu$_{2}$Se$_{2}$.
There are four peaks located
at binding energies (BEs) of 163.6, 161.5, 158.4, and 156.4 eV, which
can be attributed to Bi-$4f_{5/2}$, Y-$3d_{3/2}$, Bi-$4f_{7/2}$,
and Y-$3d_{5/2}$ respectively, indicating the Bi and Y ions adopt
the valence of +3.\cite{Debies-TP-1977,Vasquez-RP-1989} The monovalence state of Cu ion was reported in other CuCh-based compounds.\cite{Jin-SF-2012} As shown in Fig. \ref{fig2-XRD-EDS}(d), the peak position of Cu-2$p_{3/2}$ is 932.24 eV. In copper oxides, the binding energies of Cu$^{+}$ (Cu$_{2}$O) and Cu$^{2+}$ (CuO) are 932.6 eV and 933.6 eV respectively.\cite{Haber-J-1978} However, in copper selenides, the binding energies of Cu$^{+}$ (Cu$_{2}$Se) and Cu$^{2+}$ (CuSe) are 931.9 eV and 932 eV respectively.\cite{Romand-R-1978} Thus, it is difficult to determine the valence of Cu in Bi$_{2}$YO$_{4}$Cu$_{2}$Se$_{2}$ just through the measurement of XPS. Because the valences of Bi and Y are +3 and the valences of Se and O are -2 in this compound, the valence of Cu should be +1.5, i.e., a mixed valence state of Cu$^{2+}$/Cu$^{1+}$. A 3$d^{9.5}$ configuration of Cu ion in Bi$_{2}$YO$_{4}$Cu$_{2}$Se$_{2}$ can be expected. The existence of partial Cu$^{2+}$ was verified by the measurement of EPR, which will be discussed below.

\begin{table}
\caption{\label{Table2-atomic-coordinates}Fractional atomic coordinates of
Bi$_{2}$YO$_{4}$Cu$_{2}$Se$_{2}$ and BiOCuSe determined by two-phase refinement of X-ray
data. }

\begin{tabular}{cccc}
\hline
\hline
Atom & $x$ & $y$ & $z$\tabularnewline
\hline
Bi$_{2}$YO$_{4}$Cu$_{2}$ Se$_{2}$  & (wt\%: 96.76\%)  \tabularnewline
Bi & 0.5 & 0.5 & 0.8968(1)\tabularnewline
Y & 0.5 & 0.5 & 0.5\tabularnewline
O & 0 & 0.5 & 0.9353(7)\tabularnewline
Cu & 0 & 0.5 & 0.25\tabularnewline
Se & 0.5 & 0.5 & 0.3110(2)\tabularnewline
\hline
BiOCuSe  & (wt\%: 3.24\%) \tabularnewline
Bi & 0.25 & 0.25 & 0.141(2)\tabularnewline
Cu & -0.25 & 0.25 & 0.5\tabularnewline
Se & -0.25 & -0.25 & 0.342(5)\tabularnewline
O & -0.25 & 0.25 & 0\tabularnewline
\hline
\hline
\end{tabular}
\end{table}

\begin{table}

\caption{\label{Table3-Structural-parameters}Structural parameters obtained from Rietveld analysis at room temperature.}

\begin{tabular}{ccc}
\hline
\hline
 & Bi$_{2}$YO$_{4}$Cu$_{2}$ Se$_{2}$ & BiOCuSe\tabularnewline
\hline
space group & I4/mmm & P4/nmm\tabularnewline
$a$ (\AA) & 3.8628(3) & 3.9311(7)\tabularnewline
$c$ (\AA) & 24.322(2) & 8.899(3)\tabularnewline
bond length &  & \tabularnewline
$d_{\mathrm{Bi-O}}$(\AA) & 2.399(1) & 2.34(3)\tabularnewline
$d_{\mathrm{Bi-Se}}$(\AA) & 3.437(4) & 3.30(1)\tabularnewline
$d_{\mathrm{Cu-Se}}$(\AA) & 2.435(3) & 2.42(2)\tabularnewline
bond angle &  & \tabularnewline
O-Bi-O & $79.02(3)\times4$ & $73.06(1)\times4$\tabularnewline
 & $128.27(3)\times2$ & \tabularnewline
O-Y-O & $66.48(3)\times4$ & \tabularnewline
 & $101.7(4)\times2$ & \tabularnewline
Se-Cu-Se & $104.94(1)\times2$ & $108.75(1)\times2$\tabularnewline
 & $111.78(1)\times4$ & $109.84(2)\times4$\tabularnewline
\hline
\hline
\end{tabular}

\end{table}

\subsection{Physical Properties. }

\begin{figure}
\includegraphics[width=0.8\columnwidth]{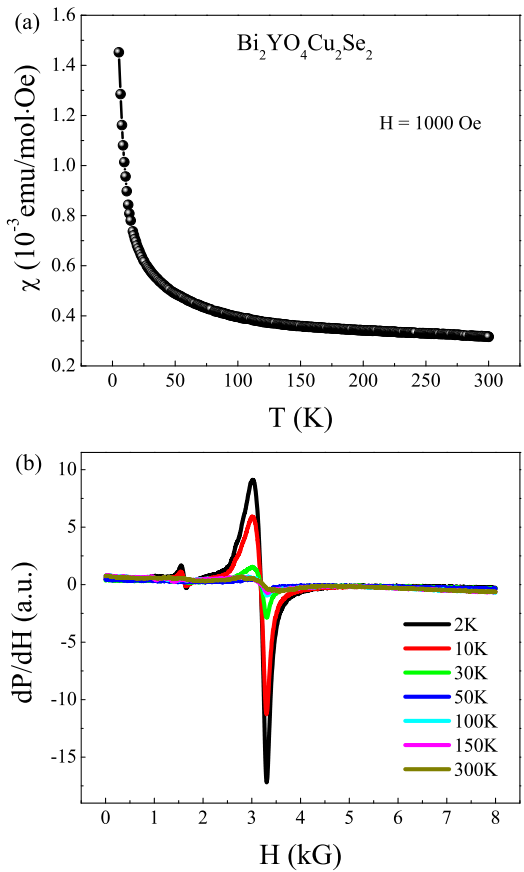}

\caption{\label{fig3-MT}(a) The temperature dependence of magnetic susceptibility $\chi(T)$ curve from 5 to 300K. (b) the EPR spectra of Bi$_{2}$YO$_{4}$Cu$_{2}$Se$_{2}$ at different temperatures.
 }

\end{figure}

Figure \ref{fig3-MT} (a) shows the temperature dependence of the magnetic susceptibility of Bi$_{2}$YO$_{4}$Cu$_{2}$Se$_{2}$. Since the magnetic signal of the sample is too weak to be distinguished from the signal of background using the routine measurement, we took a two-step measurement: Firstly, we measured the signal of the background, and then  the total signal of background and sample was measured. After subtracting the signal of the background from the raw data, a paramagnetic (PM)-like temperature dependence of susceptibility was observed with no evidence of magnetic ordering at low temperatures due to the monotonous increase in the magnetic susceptibility. Since EPR is helpful for understanding magnetic interactions and spin correlation on a microscopic level, we performed the EPR measurements. Figure \ref{fig3-MT}(b) shows the EPR spectra $dP/dH$ of Bi$_{2}$YO$_{4}$Cu$_{2}$Se$_{2}$ at 2, 10, 30, 50, 100, 150, and 300 K. There are two resonance lines centered at $\sim$ 3200 Gauss and $\sim$1600 Gauss, respectively. And both resonance fields are independent of temperature implying the lack of spin fluctuation. The Cu(I) ions ($3d^{10}$) should be EPR-inactive due to the lack of unpaired electrons. Therefore, the PM resonance signal at ~3200 Gauss corresponding to the free-electron $g$ factor $g_{e}=2$ is attributed to the existence of Cu$^{2+}$ ions, which is in agreement with the XPS results. Moreover, the weak resonance signal at ~1600 Gauss is often concomitant with the PM resonance signal at ~3200 Gauss in the Cu-contained compounds and is termed as half-field transition arising from the formation of a triplet state Cu(II) dimers.\cite{Svorec-J-2009} As a results, the existence of magnetic ordering and/or spin fluctuation in Bi$_{2}$YO$_{4}$Cu$_{2}$Se$_{2}$ can be ruled out based on the present magnetic and EPR results.

\begin{figure}
\includegraphics[width=0.8\columnwidth]{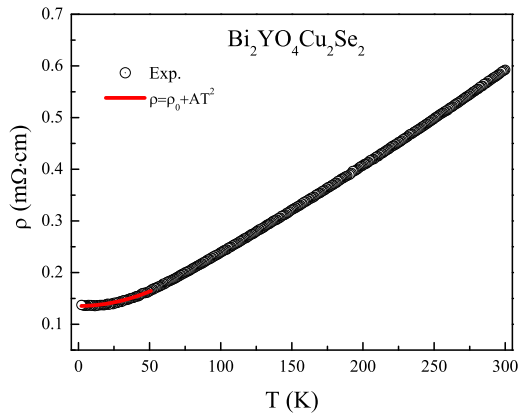}

\caption{\label{fig4-RT}Electrical resistivity of Bi$_{2}$YO$_{4}$Cu$_{2}$Se$_{2}$
is plotted as a function of temperature from 300 to 2 K, and the red
line is the Fermi liquid fitting of the resistivity at low temperature. }

\end{figure}

As shown in Fig. \ref{fig4-RT}, the resistivity $\rho(T)$ of polycrystalline
Bi$_{2}$YO$_{4}$Cu$_{2}$Se$_{2}$ shows a metallic behavior in
the measured temperature region. The resistance drops linearly with
temperature from 300 to 100 K and the room temperature value of resistivity
is $0.59\mathrm{\, m}\Omega\mathrm{\cdot cm}$, which is much smaller
than the reported value.\cite{Evans-J-S-O-2002} The temperature
coefficient of resitivity is $1.7\times10^{-3}$ m$\Omega$$\cdot$cm/K.
It should be noted that the polycrystalline BiOCuSe shows semiconducting
behavior. The impurity may have some minor influence on the absolute
value of resistivity, but the metallic behavior should be intrinsic.
As can be seen, below 45 K the resistivity is satisfied with the equation
\begin{equation}
\rho(T)=\rho_{0}+AT^{2},\label{eq:eq-rt}
\end{equation}
which is suggestive of Fermi liquid behavior in the ground state.
The residual resistivity $\rho_{0}$ and parameter $A$ are found to
be $134.7(3)\,\mu\Omega\cdot\mathrm{cm}$ and $0.01118(3)\,\mu\Omega\cdot\mathrm{cm/K^{2}}$
respectively. The metallic behavior is consistent with our theoretical
calculation results as shown below.

On the other hand, previous theoretical reports have indicated that
the undoped BiOCuSe is a semiconductor and the metallic behavior of
BiOCuSe is caused by Cu deficiencies which introduce holes to the
system.\cite{Hiramatsu-H-2008,Barreteau-C-2012} In undoped BiOCuSe,
the ``electricity storage'' layer Bi$_{2}$O$_{2}$
provides two electrons to the ``conductive''
layer Cu$_{2}$Se$_{2}$. However, in Bi$_{2}$YO$_{4}$Cu$_{2}$Se$_{2}$,
the ``charge reservoir'' layer
Bi$_{2}$YO$_{4}$ provides only one electron to the ``conductive''
layer Cu$_{2}$Se$_{2}$. Therefore, the replacement of Bi$_{2}$O$_{2}$
by Bi$_{2}$YO$_{4}$ should introduce one hole and as a result, Bi$_{2}$YO$_{4}$Cu$_{2}$Se$_{2}$
is a metal with $p$-type carriers. There will be further discussion
in the calculation part.

\begin{figure}
\includegraphics[width=0.8\columnwidth]{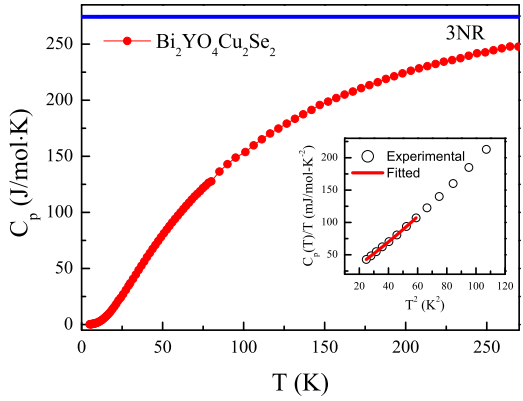}

\caption{\label{fig5-Cp}Temperature dependence of specific heat for Bi$_{2}$YO$_{4}$Cu$_{2}$Se$_{2}$.
The inset is the low temperature data $C_{p}(T)/T$ plotted as a function
of $T^{2}$. }
\end{figure}

Figure \ref{fig5-Cp} shows the specific heat of Bi$_{2}$YO$_{4}$Cu$_{2}$Se$_{2}$
measured from 280 K to 5 K. The specific heat of Bi$_{2}$YO$_{4}$Cu$_{2}$Se$_{2}$
approaches to the value of $3NR$ at 280 K, where $N$ is the atomic
number in the chemical formula ($N=11$) and $R$ is the gas constant
($R=8.314$ J/mol$\cdot$K), consistent with the Dulong-Petit law. The inset
shows the low-temperature specific heat data, $C_{p}(T)/T$ plotted
as a function of $T^{2}$. It can be fitted using
\begin{equation}
C_{p}=\gamma T+\beta T^{3}+\delta T^{5},\label{eq:eq-cp}
\end{equation}
where $\gamma T$ is the electronic contribution with $\gamma$ the
Sommerfeld coefficient, and $\beta T^{3}+\delta T^{5}$, originate
from the lattice contribution. The fitted values of $\gamma$, $\beta$,
and $\delta$ are equal to 3.6(8) mJ/mol$\cdot$K, 1.56(3) mJ/mol$\cdot$K$^{4}$
, and 0.0038(3) mJ/mol$\cdot$K$^{6}$ respectively. Using the formula
\begin{equation}
\Theta_{D}=\left(\frac{n\times1.944\times10^{6}}{\beta}\right)^{1/3},\label{eq:eq-debye}
\end{equation}
where $n$ is the number of atoms in a unit cell, we derived the Debye
temperature $\Theta_{D}=239$ K. According to the Sommerfeld theory
of metals, $\gamma$ can be written as function of the density of
states (DOS) at Fermi level as:
\begin{equation}
\gamma=\frac{\pi^{2}k_{B}^{2}}{3}N(E_{F})(1+\lambda),\label{eq:eq-gamma}
\end{equation}
where $N(E_{F})$ is the DOS at the Fermi level and $\lambda$ is
the electron-phonon coupling constant. Assuming $\lambda$ is 0, the
obtained value of $N(E_{F})$ is 1.5 states/eV/f. u..

\begin{figure}

\includegraphics[width=0.8\columnwidth]{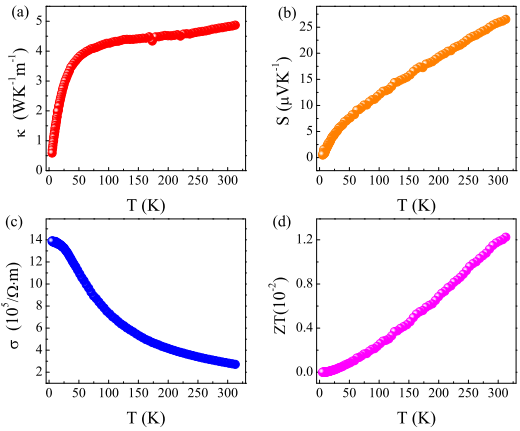}\caption{\label{fig6-Thermoelectric}(a) The thermal conductivity $\kappa$,
(b) Seebeck coefficient $S$, (c) electrical conductivity $\sigma$
and (d) figure of merit $ZT$ as a function of temperature from 330
to 5 K.}
\end{figure}
The thermal transport measurement results including the Seebeck coefficient
$S$, electrical conductivity $\sigma$, and thermal conductivity
$\kappa$ are shown in Fig. \ref{fig6-Thermoelectric}. The $S$
is positive in the whole measurement temperature region, indicating
the major carriers are holes. The $S(T)$ is nearly linearly dependent
with temperature and drops to zero at low temperature. The $\kappa(T)$
is almost constant when $T$ is above 100 K, below which the $\kappa(T)$
drops fast near to zero. The dimensionless figure of merit, $ZT$
($ZT=S^{2}T\sigma/\kappa$), which represents
for the efficiency of a thermoelectric material, was calculated and
presented in Fig. \ref{fig6-Thermoelectric} (d). Due to the low
resistivity, and relatively small thermal conductivity, the $ZT$ value
of our sample at 300 K reaches about 0.012. And the value of $ZT$
increases smoothly from 5 K to 300 K. It implies Bi$_{2}$YO$_{4}$Cu$_{2}$Se$_{2}$
may be a potential thermoelectric material at high temperatures.

\begin{figure}
\includegraphics[width=0.8\columnwidth]{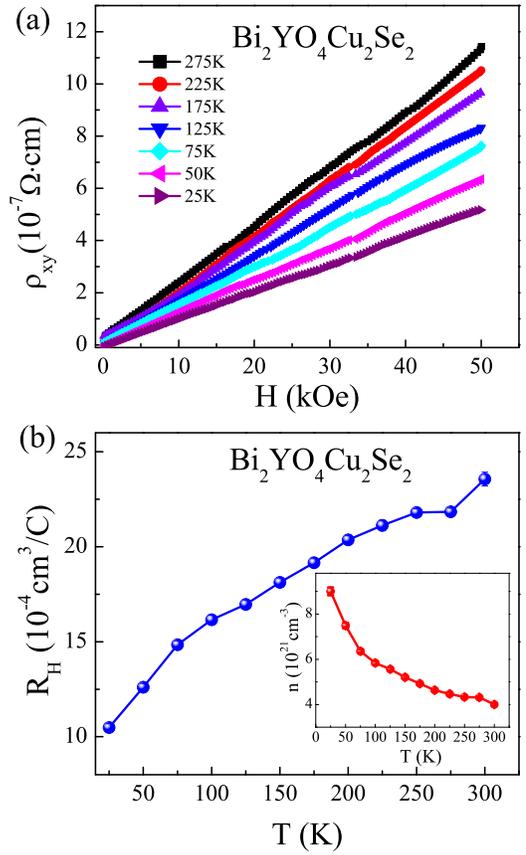}

\caption{\label{fig7-hall} (a) Field dependence of $\rho_{xy}(H)$ at various
temperatures. (b) Temperature dependence of the Hall coefficient $R_{H}$
of Bi$_{2}$YO$_{4}$Cu$_{2}$Se$_{2}$. Inset: temperature dependence
of the carrier density $n=1/|eR_{H}|$ calculated from $R_{H}$.}

\end{figure}

In order to confirm the type of carrier and determine the carrier density in Bi$_{2}$YO$_{4}$Cu$_{2}$Se$_{2}$, the magnetic field
dependence of the Hall resistivity at various temperatures was measured. (Figure \ref{fig7-hall} (a)) The transverse resistivity $\rho_{xy}(H)$ is positive at all measured temperatures, indicating that the hole-type carrier is dominant, consistent with the sign of $S(T)$. The Hall coefficient $R_{H}=\rho_{xy}(H)/H$ at different temperatures is shown in Fig. \ref{fig7-hall} (b). It can be seen that $R_{H}$ decreases with temperature. The change can be ascribed to the multiband effect, which has been observed in classic two-band superconducting materials such as MgB$_{2}$ \cite{Yang-H-2008} as well as in iron based material Nd(O,F)FeAs\cite{Cheng-P-2008}. A multiband electronic structure at the Fermi level is also supported by the DFT calculations discussed below.

\subsection{Electronic Structure of Bi$_{2}$YO$_{4}$Cu$_{2}$Se$_{2}$. }
\begin{table}
\caption{\label{Table4-optimized-lattice}The optimized lattice parameters
and atomic coordinates compared with experimental values.}

\begin{tabular}{cccccc}
\hline
\hline
 &  & Exp. &  & Opt. & \tabularnewline
\hline
$a$(\AA) & ~~~~~~~~~~ & 3.8628(3) & ~~~~~ & 3.888 & \tabularnewline
$c$(\AA) & ~ & 24.322(2) & ~ & 24.872 & \tabularnewline
$z$(Bi) & ~ & 0.8968(1) & ~ & 0.89895 & \tabularnewline
$z$(Y) & ~ & 0.5 & ~ & 0.5 & \tabularnewline
$z$(O) & ~ & 0.9353(7) & ~ & 0.94445 & \tabularnewline
$z$(Cu) & ~ & 0.25 & ~ & 0.25 & \tabularnewline
$z$(Se) & ~ & 0.3110(2) & ~~ & 0.31106 & \tabularnewline
\hline
\hline
\end{tabular}

\end{table}

The crystal structure of Bi$_{2}$YO$_{4}$Cu$_{2}$Se$_{2}$ was optimized with respect to lattice parameters and atomic positions. The optimized lattice parameters and atomic coordinates are listed in Table \ref{Table4-optimized-lattice}. The structure is in good agreement with experimental observation except for the slight overestimation of $c$-axial lattice parameter (($c_{opt.}$-$c_{exp.}$)/$c_{exp.}$ = 2.3 \%), which is expected within the GGA calculation. Nonmagnetic (NM), ferromagnetic (FM), checkerboard and stripes antiferromagnetic (AFM) states are tested in the system. The magnetic moments of each atom in FM and AFM states are converged to zero, which is consistent with the magnetic properties measured in experiment.

\begin{figure}
\includegraphics[width=0.9\columnwidth]{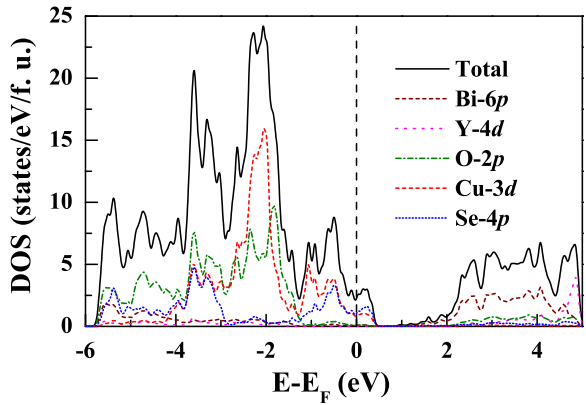}

\caption{\label{fig8-dos} The calculated DOS of Bi$_{2}$YO$_{4}$Cu$_{2}$Se$_{2}$. }

\end{figure}

Figure \ref{fig8-dos} shows the calculated DOS of Bi$_{2}$YO$_{4}$Cu$_{2}$Se$_{2}$. The finite DOS at Fermi level ($E_{F}$) indicates the metallic ground state, consistent with the experimental results. In the valence band, Bi-$6p$ electrons hybridize with O-$2p$ electrons in the energy range between -6 eV and -1 eV, and the hybridization of Cu-$3d$ and Se-$4p$ electrons is from -6 eV to 0.5 eV across $E_{F}$. The conduction bands at higher energy, mainly built up of Bi-$6p$ and O-$2p$ electrons, are separated by a small gap of ~0.25 eV. The $E_{F}$ locates at a small valley of DOS, leading to the small $N(E_{F})$. The calculated $N(E_{F})$ (2.1
states/eV/f. u.) is slightly larger than the experimental one (1.5 states/eV/f. u.), which may be due to existence of the small amount of Se deficiency in our polycrystalline sample and the semiconducting impurity BiOCuSe.

\begin{figure*}

\includegraphics[width=0.6\textwidth]{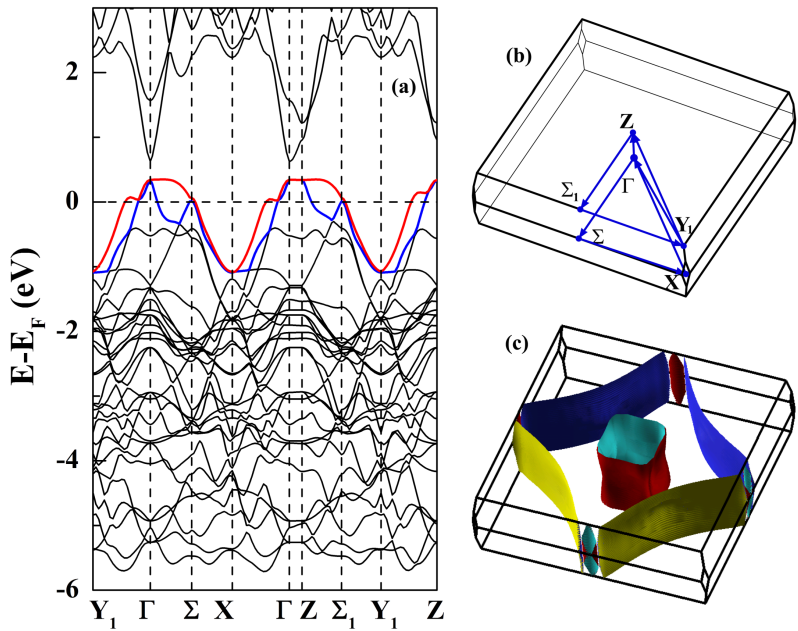}\caption{\label{fig-9-band} (a) The band structure, (b) The Brillouin zone and (c)Fermi surface of Bi$_{2}$YO$_{4}$Cu$_{2}$Se$_{2}$.
The red and blue lines in (a) denote the bands crossing E$_{F}$. The high symmetry points are denoted in (b).}
\end{figure*}

Figure \ref{fig-9-band}(a) shows the band structure of Bi$_{2}$YO$_{4}$Cu$_{2}$Se$_{2}$. In the $\Gamma$-Z direction (perpendicular to the layers), the bands are rarely dispersive. On the other hand, the strong hybridizations between Cu and Se give rise to a strongly dispersive band structure in the layer-parallel directions. Similar quasi-2D properties are found in alkali metal intercalated compound KCu$_{2}$Se$_{2}$, which also behaves metallic.\cite{Tiedje-O-2003} There are two bands crossing $E_{F}$. For the lower one (line in blue in Fig. \ref{fig-9-band}(a)), there are two hole cylinders: One locates at zone center, and another locates around $\Sigma$ and $\Sigma$$_{1}$ points. The large electron cylinder around the zone corner is from the upper band crossing $E_{F}$ (line in red in Fig. \ref{fig-9-band} (a)). The total volume outside the electron cylinders and the volume confined in the hole cylinders contain one hole/f. u..

In order to interpret the temperature dependence of hall coefficient qualitatively, the simple two-band model was introduced\cite{Cheng-P-2008,Mu-G-2010}:
\begin{equation}
R_{H}=\frac{\sigma_{1}^{2}R_{1}+\sigma_{2}^{2}R_{2}}{(\sigma_{1}+\sigma_{2})^{2}},\label{eq:eq-hall}
\end{equation}
where $R_{i}=1/n_{i}q$ represents the Hall coefficient for the carriers in each band separately with $n_{i}$ the concentration of the charge carriers in the different bands. $\sigma_{i}$ (i = 1,2) is the conductance of the charge carriers in different bands, which can be expressed as $\sigma_{i}=\frac{|q^{2}|\tau_{i}n_{i}}{m_{i}^{*}}$, where the $m_{i}^{*}$ is the effective mass, and $\tau_{i}$ is the relaxation time in each band. Normally, the carrier concentration rarely varies with the temperature. Since the multiple Fermi pockets locate at different position with different shape in Bi$_{2}$YO$_{4}$Cu$_{2}$Se$_{2}$, it is natural that they have different effective mass and Fermi velocity of them. The scattering should also share the low similarities. Therefore, the carriers in different bands may have different $\tau_{i}$, which varies differently with temperature. Each band can produce complex contributions to the total $R_{H}$, which lead to the variation of $R_{H}$ with temperature.

\begin{figure}

\includegraphics[width=0.99\columnwidth]{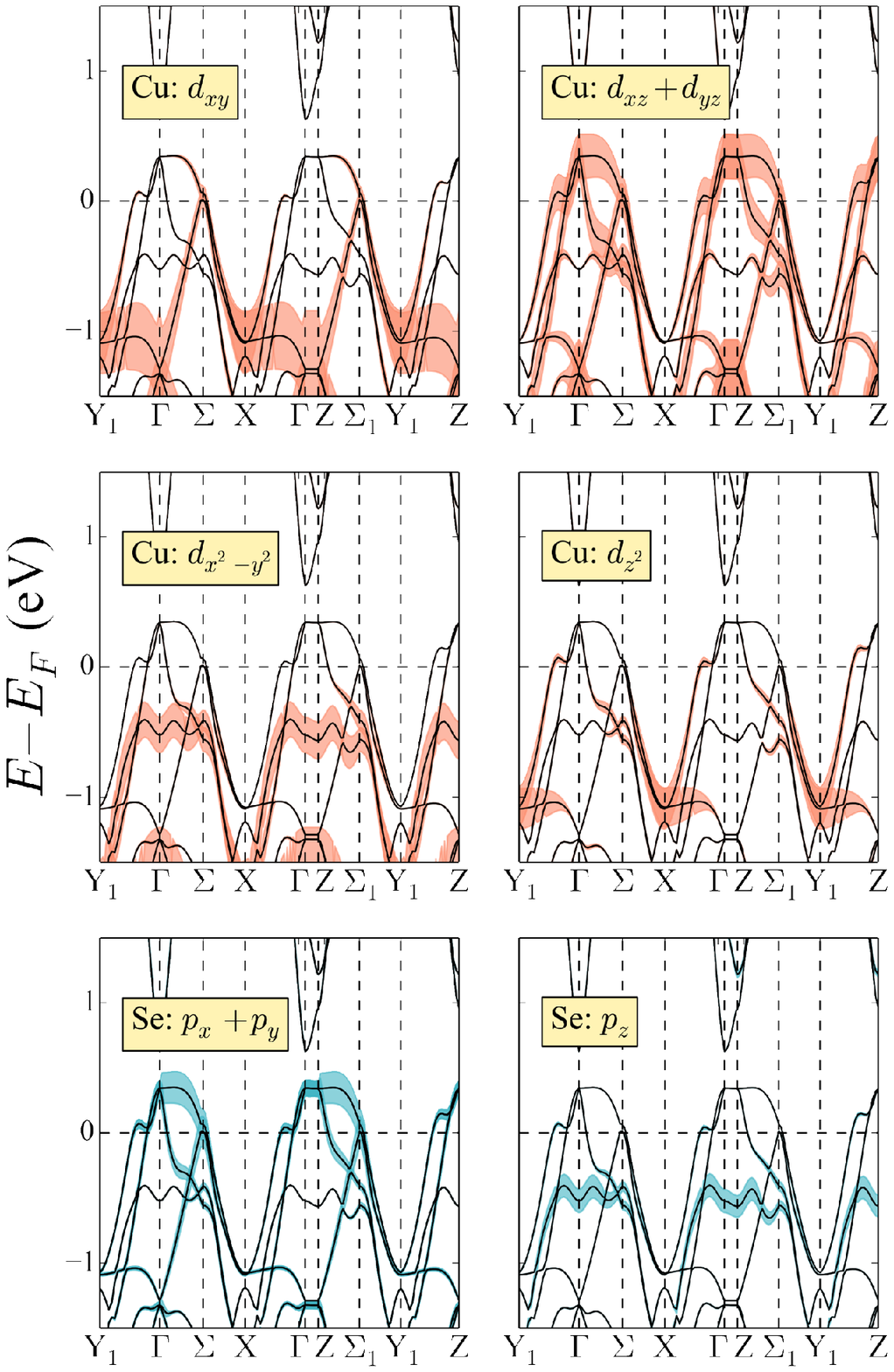}\caption{\label{fig10-fatband}\textquotedblleft{}Fat bands\textquotedblright{}
of Bi$_{2}$YO$_{4}$Cu$_{2}$Se$_{2}$, decorated with partial orbital
characters of Cu and Se.}

\end{figure}

In order to analyze the orbital characters, the \textquotedblleft{}fat-band\textquotedblright{}
of Bi$_{2}$YO$_{4}$Cu$_{2}$Se$_{2}$ is also presented (Fig. \ref{fig10-fatband}). Obviously,
the Fermi surface is formed by the bands with prominent hybridization
between $d_{xy}+d_{yz}$ of Cu and $p_{x}+p_{y}$ of Se. The Se-$p_{z}$
orbital has almost no contribution at $E_{F}$, indicating the absence
of conduction between the layers. Such quasi-2D electronic properties
of Bi$_{2}$YO$_{4}$Cu$_{2}$Se$_{2}$ needs to be be further confirmed
by experimental studies on the single crystal.

\begin{figure}
\includegraphics[width=0.95\columnwidth]{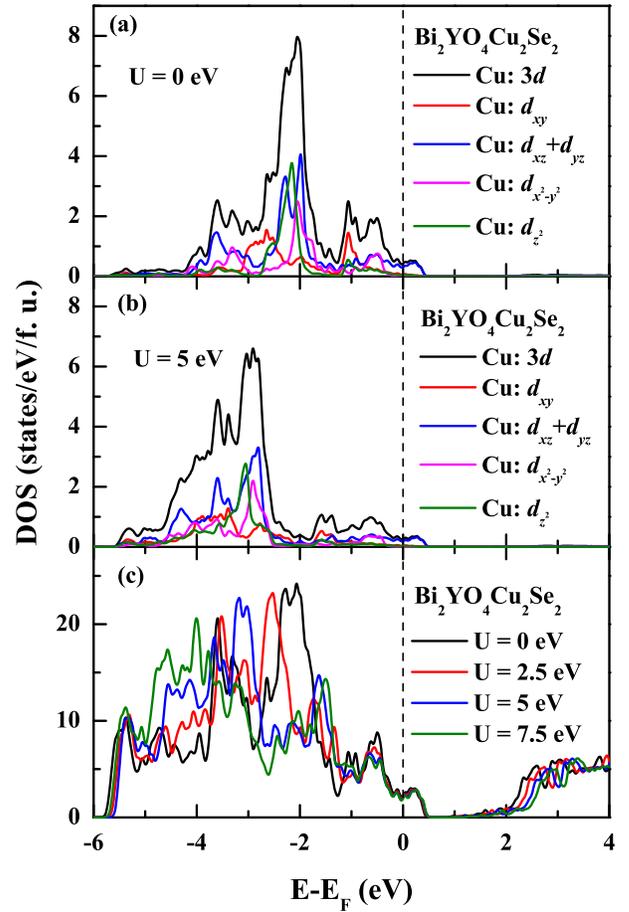}\caption{\label{fig11-Udos} (a) Orbital contribution of DOS of Cu-$3d$ electrons
with $U=0$ eV. (b) Orbital contribution of DOS of Cu-$3d$ electrons
with $U=5$ eV. (c) DOS of Bi$_{2}$YO$_{4}$Cu$_{2}$Se$_{2}$ with
$U=$ 0, 2.5, 5, and 7.5 eV.}

\end{figure}

We also permormed a GGA+$U$ calculation for NM, FM, checkerboard AFM, and stripes AFM states
considering the strong Coulomb
correlation of $3d$ electrons in Bi$_{2}$YO$_{4}$Cu$_{2}$Se$_{2}$. The magnetic moments of each atom in FM and AFM states are converged to zero as well.
The corresponding DOS of NM state are presented in Fig. \ref{fig11-Udos}. It
can be found that the Coulomb interactions moves the weight of $3d$
electrons towards lower energy. Such properties have been reported
in some NiSe-based superconductors.\cite{Lu-F-2012} Under different
$U$, the $d_{xy}+d_{yz}$ orbitals remain contributing prominently
at $E_{F}$, which makes the $N(E_{F})$ rarely varies. Unquestionably, the added Coulomb interactions do not make the system become a Mott insulator. It might be because of the $3d^{9.5}$ configuration of Cu in Bi$_{2}$YO$_{4}$Cu$_{2}$Se$_{2}$, for which the five orbitals of Cu-$3d$ are almost full filled.\cite{Lu-F-2012}

\subsection{Ground States of Bi$_{2}$YO$_{4}$Cu$_{2}$Se$_{2}$-type Compounds:
A DFT Analysis}

\begin{figure}
\includegraphics[width=0.95\columnwidth]{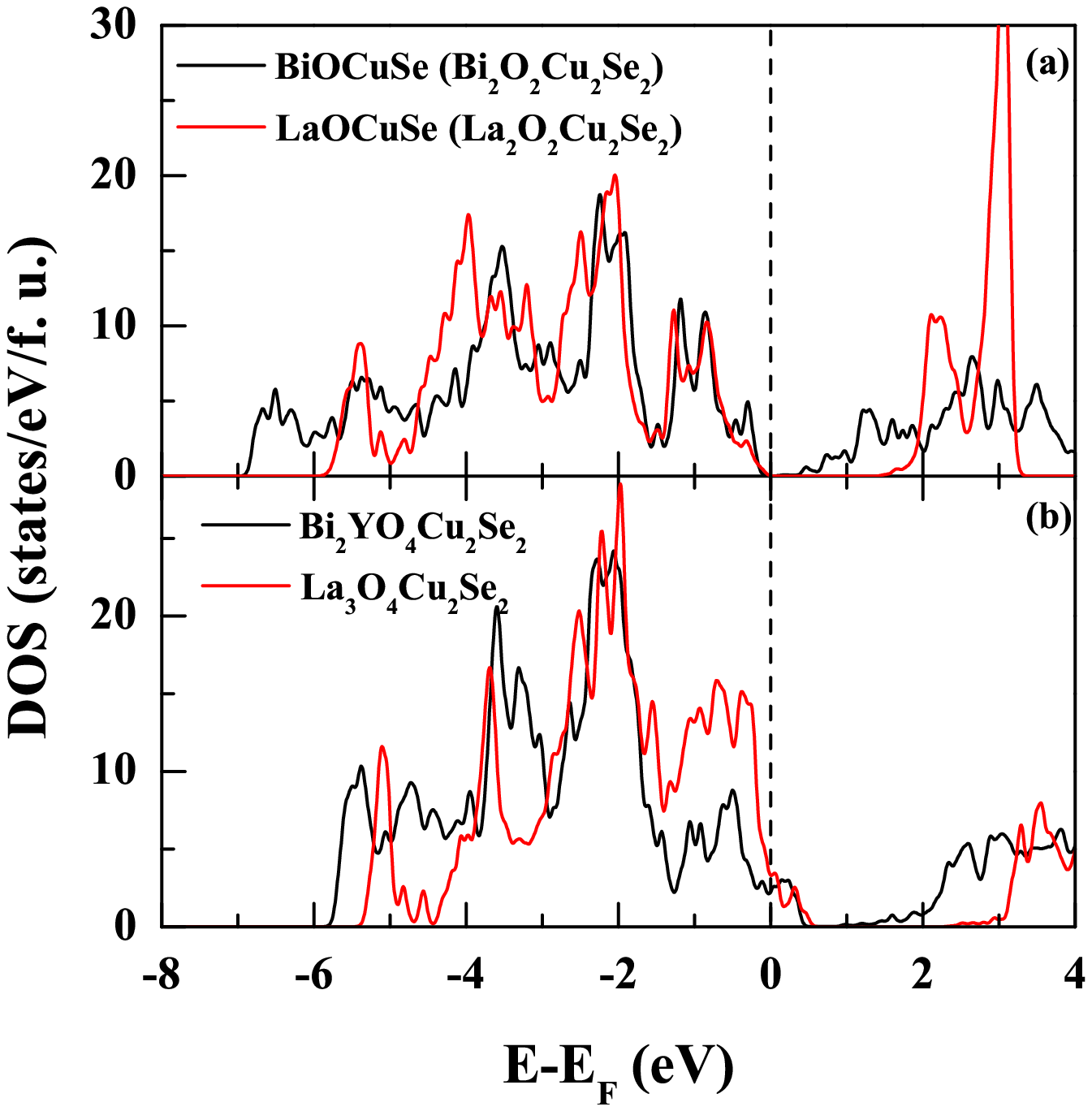}

\caption{\label{fig12-1111} (a) The DOS of BiOCuSe and LaOCuSe. (b) The DOS
of Bi$_{2}$YO$_{4}$Cu$_{2}$Se$_{2}$, and a hypothetical La$_{3}$O$_{4}$Cu$_{2}$Se$_{2}$.}
\end{figure}

Based on the calculation above, the metallic ground state of Bi$_{2}$YO$_{4}$Cu$_{2}$Se$_{2}$
is verified. However, to the best of our knowledge, most undoped CuCh(Ch=S,
Se)-based compounds are semiconductors (see Table \ref{Table1-konwn-CuSe}).
The origin of metallic ground states of Bi$_{2}$YO$_{4}$Cu$_{2}$Se$_{2}$-type compounds
need to be distinctly clarified. Generally speaking, CuSe-based compounds
can be seen as Cu$_{2}$Se$_{2}$ layer intercalated by some atoms
or molecules. Since the valence band maximum (VBM) is mainly composed
of the Cu-Se hybridization states, the function of intercalated blocking
layers can be simply seen as giving their valence electrons to the
bands. All the reported CuSe-based semiconductors have the bivalent blocking layers, in which Cu-$3d$ bands are full filled. When the valence of the blocking layer is smaller than +2, the Cu-$3d$ bands are partly occupied, leading to the metallic ground state (e.g. [HgO]$^{0+}$CuSe\cite{Kim-GC-2012} and K$^{1+}$Cu$_{2}$Se$_{2}$\cite{Tiedje-O-2003}).
Figure \ref{fig12-1111} shows the DOS of BiOCuSe, LaOCuSe, Bi$_{2}$YO$_{4}$Cu$_{2}$Se$_{2}$,
and hypothetical La$_{3}$O$_{4}$Cu$_{2}$Se$_{2}$.
For the semiconducting BiOCuSe and LaOCuSe, the Bi$_{2}$O$_{2}$/La$_{2}$O$_{2}$
blocking layers are bivalent. According to the XPS
measurements mentioned above, Bi$_{2}$YO$_{4}$ layer only gives
one valence electron. Similarly, the La$_{3}$O$_{4}$ layer should
be monovalent. In that case, there is one hole in the valence band,
making the systems become metals with $p$-type carriers. Such trend
is consistent with the case of ACu$_{2}$Se$_{2}$ (A is alkali or
alkali earth metal).\cite{Tiedje-O-2003}

\begin{figure}
\includegraphics[width=0.95\columnwidth]{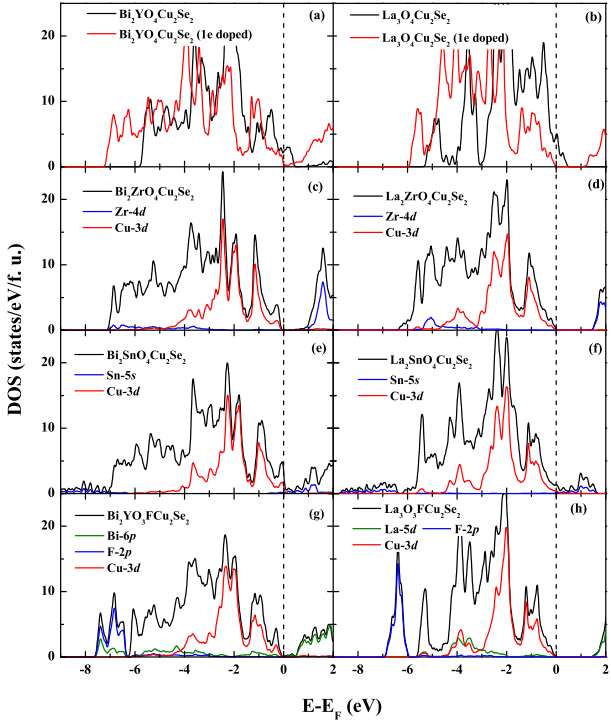}\caption{\label{fig13-doped} The DOS of one electron doped in Bi$_{2}$YO$_{4}$Cu$_{2}$Se$_{2}$
(left panel) and La$_{3}$O$_{4}$Cu$_{2}$Se$_{2}$ (right panel).
(a) and (b) show the doping effect simulated by simply adding a electron
to the system based on RBA. (c)-(h) show the doping effect simulated
by doping typical tetravalent cations Zr and Sn or a typical monovalent
anion F. }
\end{figure}

If this trend is universal for all the CuSe-based compounds, one may
expect that the one electron doping will make Bi$_{2}$YO$_{4}$Cu$_{2}$Se$_{2}$/La$_{3}$O$_{4}$Cu$_{2}$Se$_{2}$
semiconducting. We calculated the DOS under such doping of Bi$_{2}$YO$_{4}$Cu$_{2}$Se$_{2}$
and La$_{3}$O$_{4}$Cu$_{2}$Se$_{2}$ using the rigid band approximation
(RBA) (Fig. \ref{fig13-doped} (a) and (b)). The results show that
the electron doping raises $E_{F}$ to the VBM. In reality, such doping
needs to be processed by replacing the trivalent cation Bi/Y/La with
a tetravalent cation, or replacing the bivalent anion O with monovalent
anions. So we calculated the DOS of Bi$_{2}$YO$_{4}$Cu$_{2}$Se$_{2}$
and La$_{3}$O$_{4}$Cu$_{2}$Se$_{2}$ with the substitution of the
typical tetravalent cations Zr and Sn, and monovalent anion F. The optimized structural parameters of these doped samples are presented in Table \ref{Table5-optimized lattice parameters}.  Figures \ref{fig13-doped} (c)-(h) show the calculated DOS of these samples. The $E_{F}$ is indeed moved to the top of Cu-3$d$ bands by the substitutions. For Bi$_{2}$ZrO$_{4}$Cu$_{2}$Se$_{2}$, La$_{2}$ZrO$_{4}$Cu$_{2}$Se$_{2}$, and La$_{3}$O$_{3}$FCu$_{2}$Se$_{2}$, the semiconducting ground states are obtained as expected. Surprisingly, for the Sn-doped samples, the bands of Sn-5$s$ occur at $E_{F}$, leading to the metallic ground state. For Bi$_{2}$YO$_{3}$FCu$_{2}$Se$_{2}$, the hybridizations between Bi and F move the conduction band downwards, and as a result, a pseudogap was obtained. Obviously, when the Cu-3$d$ bands are fully occupied, the bands of the blocking layer determine the ground states for these samples.

\begin{table}

\caption{\label{Table5-optimized lattice parameters}The optimized lattice parameters of some hypothetical compounds.}

\begin{tabular}{ccccccc}
\hline
\hline
  & ~ & a(\AA)& & b(\AA) & & c(\AA)\tabularnewline
\hline
Bi$_{2}$ZrO$_{4}$Cu$_{2}$Se$_{2}$ &~ & 3.792 & ~ ~& 3.792 & ~ ~& 23.983\tabularnewline
Bi$_{2}$SnO$_{4}$Cu$_{2}$Se$_{2}$ & ~ &3.832 & ~ & 3.832 & ~ & 23.978\tabularnewline
Bi$_{2}$YO$_{3}$FCu$_{2}$Se$_{2}$ & ~ & 3.845 &~& 3.953 &~& 23.998\tabularnewline
La$_{3}$O$_{4}$Cu$_{2}$Se$_{2}$ & ~ & 4.091 & ~ & 4.091 & ~ & 24.258\tabularnewline
La$_{2}$ZrO$_{4}$Cu$_{2}$Se$_{2}$ & ~ & 3.897 & ~ & 3.897 & ~ & 23.577\tabularnewline
La$_{2}$SnO$_{4}$Cu$_{2}$Se$_{2}$ & ~ & 3.938 & ~ & 3.938 & ~ & 23.409\tabularnewline
La$_{3}$O$_{3}$FCu$_{2}$Se$_{2}$ & ~ & 4.030 & ~ & 4.115 & ~ & 23.570\tabularnewline
SrFCuSe & ~ & 4.084 & ~ & 4.084 & ~ & 8.881\tabularnewline
Sr$_{3}$F$_{4}$Cu$_{2}$Se$_{2}$ & ~ & 4.073 & ~ & 4.073 & ~ & 23.752\tabularnewline
\hline
\hline
\end{tabular}

\end{table}

\begin{figure}
\includegraphics[width=0.9\columnwidth]{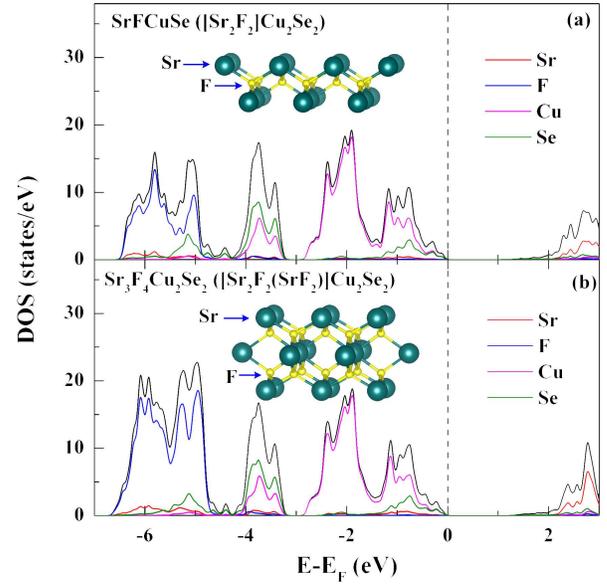}\caption{\label{fig14-SrFCuSe} The DOS of [Sr$_{2}$F$_{2}$(SF$_{2}$)$_{n}$]Cu$_{2}$Se$_{2}$ with (a) n=0 and (b) n=1. The insets show the structures of the [Sr$_{2}$F$_{2}$(SF$_{2}$)$_{n}$]-type blocking layers. }
\end{figure}

In fact, the Bi$_{2}$YO$_{4}$-type layer can be seen as a stacking of two Bi$_{2}$O$_{2}$-type slabs. If we insert more Bi$_{2}$O$_{2}$-type slabs into the compounds, we can get a series of compounds with the general formula [A$_{2}$X$_{2}$(AX$_{2}$)$_{n}$]Cu$_{2}$Se$_{2}$ (A is metal cation, X is anion). One can always choose the appropriate atoms to construct the [A$_{2}$X$_{2}$(AX$_{2}$)$_{n}$]-type blocking layer by forcing the [A$_{2}$X$_{2}$(AX$_{2}$)$_{n}$] to behave an expected valence. Figure \ref{fig14-SrFCuSe} show the semiconducting examples of [Sr$_{2}$F$_{2}$(SrF$_{2}$)$_{n}$]Cu$_{2}$Se$_{2}$ with $n=0$ and 1. The [Sr$_{2}$F$_{2}$(SrF$_{2}$)$_{n}$]-type blocking layer is bivalent no matter how many the Sr$_{2}$F$_{2}$-type slabs are introduced.

In the [A$_{2}$X$_{2}$(AX$_{2}$)$_{n}$]Cu$_{2}$Se$_{2}$-type compounds, the conduction band minimum (CBM) of such compounds are composed of the bands of cation A. When the anion X = O, in order to have a bivalent blocking layer, the valence of the cation A must be greater than or equal to +3. Some A with high valence state can lead to very deep CBM. For example, the deep CBM of Bi\cite{Hiramatsu-H-2008} leads to the small gap of Bi$_{2}$ZrO$_{4}$Cu$_{2}$Se$_{2}$ and the pesudogap of Bi$_{2}$YO$_{3}$FCu$_{2}$Se$_{2}$. The deeper CBM of Sn is overlapped with VBM, leading to the metallic ground states of Bi$_{2}$SnO$_{4}$Cu$_{2}$Se$_{2}$ and La$_{2}$SnO$_{4}$Cu$_{2}$Se$_{2}$. Therefore, the ground state of [A$_{2}$X$_{2}$(AX$_{2}$)$_{n}$]Cu$_{2}$Se$_{2}$-type compounds can be tuned by designing the blocking layers. The small/large gap semiconductors and metals can be obtained for different applications.

\section{Conclusion}

In summary, we systematically studied the physical properties of layered
oxychalcogenide Bi$_{2}$YO$_{4}$Cu$_{2}$Se$_{2}$. It crystallizes
in an unusual intergrowth structure with Cu$_{2}$Se$_{2}$ and Bi$_{2}$YO$_{4}$
layers. Bi$_{2}$YO$_{4}$Cu$_{2}$Se$_{2}$ exhibits metallic behavior
with $p$-type carriers between 2 and 300 K. We obtained a rather
large value of $ZT$ at room temperature. Theoretical calculation
confirms the quasi-2D metallic behavior of Bi$_{2}$YO$_{4}$Cu$_{2}$Se$_{2}$
and indicates the state at Fermi energy originates mainly from
Cu-$3d$ and Se-$4p$ electrons.

Based on the comparison of the transport properties of the reported
CuSe-based layered compounds, we found that the ground states of the
known CuSe-based compounds are related to the valence electrons of
their blocking layer: When the number of transferred valence electrons
from blocking layer to Cu$_{2}$Se$_{2}$ layer is smaller than two
per layer, the compounds would exhibit metallic behaviors (e.g. KCu$_{2}$Se$_{2}$\cite{Tiedje-O-2003},
HgOCuSe\cite{Kim-GC-2012}), but if that number is two, they would become semiconductors.
(e.g. BiOCuSe\cite{Hiramatsu-H-2008}). However, according to our calculations, some Bi$_{2}$YO$_{4}$Cu$_{2}$Se$_{2}$-type compounds may not obey the rules. The theoretical investigation of doping effects in the system indicates that the ground states of the Bi$_{2}$YO$_{4}$Cu$_{2}$Se$_{2}$-type compounds can be tuned by designing the blocking layers. The small/large gap semiconductors and metals can be obtained for different applications. The physical properties of other CuSe-based layered compounds need to be studied in the future in order to examine our deduction.

$Note$: In the first version, we mistyped the lattice parameters for some sample when calculating the doping effects of Bi$_{2}$YO$_{4}$Cu$_{2}$Se$_{2}$ and La$_{3}$O$_{4}$Cu$_{2}$Se$_{2}$, which influence the obtained ground states, but will not change the discussions of the properties of Bi$_{2}$YO$_{4}$Cu$_{2}$Se$_{2}$. The errors have been corrected in this version. We are sorry for such confusions.

\begin{acknowledgments}

This work was supported by the National Key Basic Research under contract No. 2011CBA00111, and the National Nature Science Foundation of China under contract Nos. 51102240, 11104279 and the Joint Funds of the National Nature Science Foundation of China and the Chinese Academy of Sciences Large-scale Scientific Facility (Grant No.U1232139) and Director's Fund of Hefei Institutes of Physical Science, Chinese Academy of Sciences. The calculations were partially performed at the Center for Computational Science, CASHIPS.

\end{acknowledgments}

\end{document}